\documentclass[12pt,a4paper]{article}

\usepackage{epsf}
\usepackage{latexsym,amssymb,euscript}
\usepackage[dvips]{graphicx}
\usepackage[numbers,sort&compress]{natbib}
\usepackage{amsmath}
\usepackage{slashed}
\usepackage{booktabs}
\usepackage{hyperref}
\usepackage{braket}
\usepackage{chngcntr}
\usepackage{bbold}
\usepackage{graphics}
\usepackage{graphicx}
\usepackage{caption}
\usepackage{subcaption}
\usepackage{pdfpages}
\usepackage[titletoc]{appendix}
\graphicspath{{./figures/}}
\hypersetup{
 linktocpage = true,
 linkcolor = blue,
 urlcolor = blue,
 colorlinks = true,
 linkcolor = teal,
 anchorcolor = blue,
 citecolor = teal,
 urlcolor = teal,
 pdfstartview = {XYZ null null 1.25} 
           }
\usepackage[left=3cm, right=2cm]{geometry}
\usepackage{pstricks}
\usepackage{color}
\usepackage{float}

\newcommand{\ddagfootnote}[1]{
\let\oldthefootnote=\thefootnote
\renewcommand{\thefootnote}{
\hspace{-0.55cm} $^{\ddagger}$}
\footnote{#1}
\let\thefootnote=\oldthefootnote
}

\newcommand{\drv}{{\rm d}}

\begin{document}

\begin{titlepage}

\begin{center}
  {\LARGE \bf Unraveling the unintegrated
  gluon distribution in the proton
  via $\rho$-meson leptoproduction}
\end{center}

\vskip 1.25cm

\centerline{\large \bf
Francesco~Giovanni~Celiberto~\hyperlink{author_mail}{\ddagfootnote{
{\it e-mail}:
\href{mailto:francescogiovanni.celiberto@unipv.it}{\hypertarget{author_mail}{francescogiovanni.celiberto@unipv.it}
}}
}}

\setcounter{footnote}{0}

\vskip .6cm


\centerline{{\sl Dipartimento di Fisica, Universit\`a degli Studi di Pavia, I-27100 Pavia, Italy}}
\vskip .2cm
\centerline{{\sl INFN, Sezione di Pavia, I-27100 Pavia, Italy}}

\vskip 2cm

\begin{abstract}
Sufficiently inclusive processes, like  the deep inelastic scattering (DIS), are described in terms of scale-dependent parton distributions, which correspond to the density of partons with a given longitudinal momentum fraction, integrated over the parton transverse momentum. For less inclusive processes, one needs to consider densities unintegrated over the transverse momentum. This work focuses on the unintegrated gluon distribution (UGD), describing the probability that a gluon can be emitted by a colliding proton, with definite longitudinal fraction and transverse momentum. Through the leptoproduction of the $\rho$-meson at HERA, existent models for the UGD will be investigated and compared with experimental data.
\end{abstract}

\vskip .5cm


\end{titlepage}

\section{Introduction}
\label{introd}

Semi-hard processes~\cite{Gribov:1984tu} (see Ref.~\cite{Celiberto:2017ius} for applications) serve as a special testing ground for calculations of high-energy scatterings in perturbative Quantum Chromodynamics (QCD). In this kinematic limit, the enhanced effect of energy logarithms compensates the smallness of the QCD coupling constant, $\alpha_s$, thus calling for an all-order resummation procedure. The most natural language to describe the resummation of these large logarithms, both in the leading (LLA) and the next-to-leading (NLA) approximation, is elegantly embodied by the Balitsky--Fadin--Kuraev--Lipatov (BFKL)~\cite{BFKL} approach. 

In the last years, a constantly increasing number of semi-hard reactions has been proposed as probe of the high-energy regime, namely: the diffractive leptoproduction of two light vector mesons~\cite{Ivanov:2004pp,Ivanov:2005gn,Ivanov:2006gt,Enberg:2005eq}, the inclusive hadroproduction of two jets featuring high transverse momenta and large separation in rapidity (better known as Mueller--Navelet process~\cite{Mueller:1986ey}), for which a richness of theoretical predictions have appeared so far~\cite{Colferai:2010wu,Caporale:2012ih,Ducloue:2013wmi,Ducloue:2013bva,Caporale:2013uva,Ducloue:2014koa,Caporale:2014gpa,Ducloue:2015jba,Caporale:2015uva,Celiberto:2015yba,Celiberto:2015mpa,Celiberto:2016ygs,Caporale:2018qnm}, the inclusive detection of two identified, light charged hadrons~\cite{Ivanov:2012iv,Celiberto:2016hae,Celiberto:2017ptm}, the multi-jet hadroproduction~\cite{Caporale:2015vya,Caporale:2015int,Caporale:2016soq,Caporale:2016xku,Celiberto:2016vhn,Caporale:2016pqe,Caporale:2016zkc}, the heavy-quark pair photo-~\cite{Celiberto:2017nyx} and hadroproduction~\cite{Bolognino:2019yls}, and more recently, $J/\Psi$-jet~\cite{Boussarie:2017oae}, hadron-jet~\cite{Bolognino:2018oth,Bolognino:2019yqj} and Drell--Yan-jet correlations~\cite{Golec-Biernat:2018kem,Deak:2018obv}. All these channels belong to a peculiar subclass of processes, where two final-state objects, well separated in rapidity, are always detected in the fragmentation region of the corresponding incoming parent particles (photon or hadron), together, in the inclusive configuration,  with an undetected gluon system, and accompanied, in the specific case multi-jet production, by the tag of one or two extra jets in more central ranges of rapidity.

\begin{figure}[tb]
\centering

\includegraphics[scale=0.50,clip]{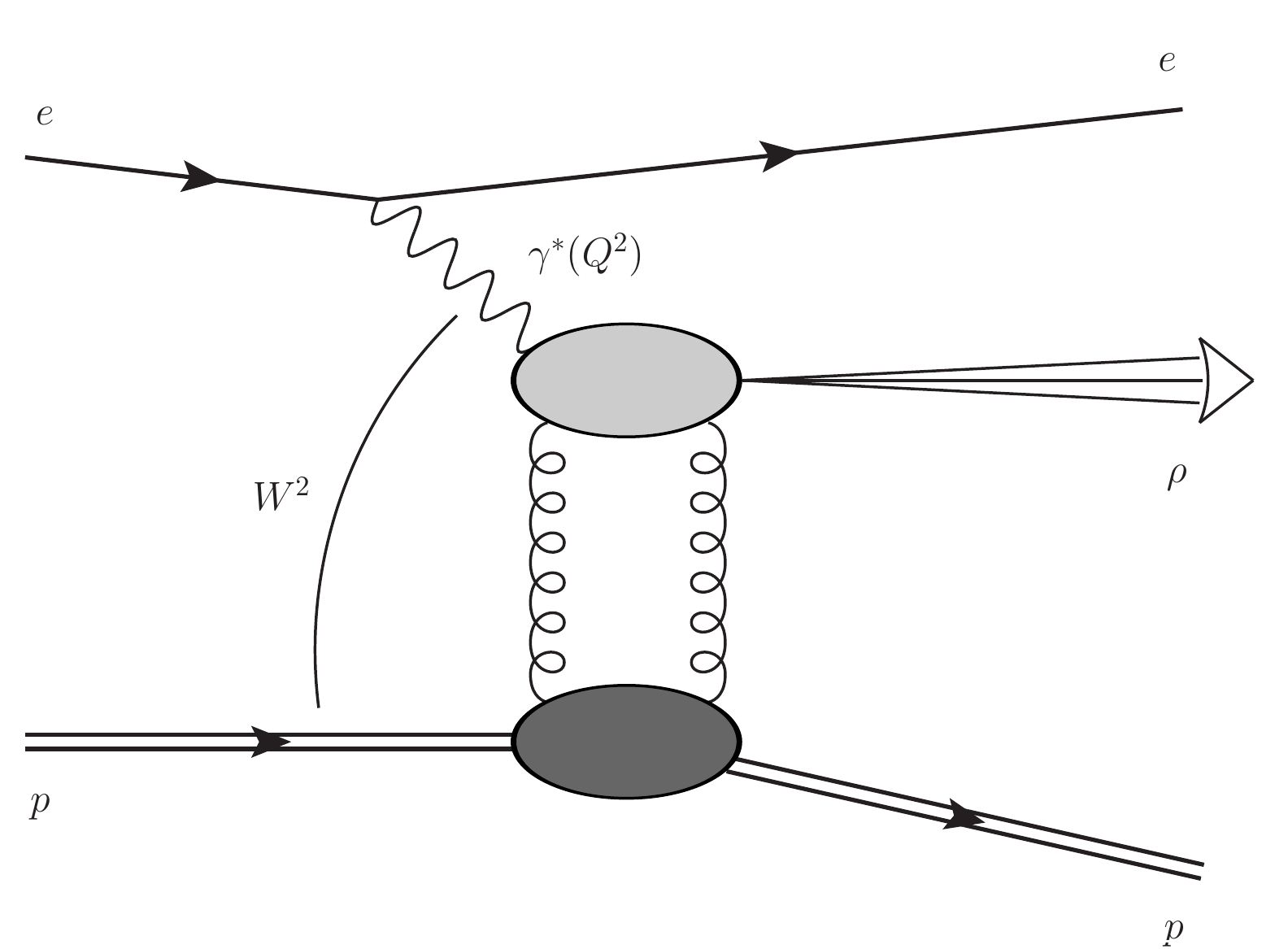}

\caption{Diagrammatic representation of the exclusive $\rho$-meson leptoproduction.}
\label{fig:process}
\end{figure}

Another interesting family of semi-hard reactions consists in the ones characterised by the emission of a single forward particle in lepton-proton collisions, as in Fig.~\ref{fig:process}. In this particular configuration it is possible to write the expression for the forward-scattering amplitude as a suitable convolution of an impact factor, describing the emission of the leptoproduced, final-state particle, and the unintegrated gluon distribution (UGD)
in the proton, which is a nonperturbative density, function of $x$ and $\kappa$, where the latter represents the gluon momentum transverse to the direction of the proton. 
This scheme is known as {\em high-energy factorisation}\footnote{An alternative and engaging formalism, formulated in the transverse-coordinate space and especially suitable to account for nonlinear evolution and gluon saturation effects, is the so-called \emph{colour dipole picture}. Interesting developments on vector meson production based on this formalism can be found in Refs.~\cite{Besse:2012ia,Besse:2013muy}.}.
The UGD, in its original definition, obeys the
BFKL~\cite{BFKL} evolution equation in the $x$ variable. Being a nonperturbative quantity, the UGD is not well known and several models for it, which lead to very different shapes in the $(x,\kappa)$-plane, have been proposed so far (see, for instance, Refs.~\cite{small_x_WG,Angeles-Martinez:2015sea}).

We show evidence~\cite{Bolognino:2018rhb,Bolognino:2018mlw,Bolognino:2019bko} that it is possible to constrain the $\kappa$-dependence of the UGD via the comparison with HERA data on helicity-dependent observables, more in depth the {\em ratio} of the two leading amplitudes for the forward polarised leptoproduction of $\rho$ mesons (Fig.~\ref{fig:process}).

\section{Theoretical setup}
\label{theory}

Widespread studies on the helicity structure of the exclusive production of $\rho$ mesons in electron-proton scattering have been conducted by the Z1 and H1 esperiments via the analysis of the subprocess (see Fig.~\ref{fig:process}):
\begin{equation}
\label{process}
\gamma^* ( \lambda_\gamma ) p \rightarrow \rho  ( \lambda_\rho ) p \, ,
\end{equation}
the meson and photon helicities, $\lambda_\rho$ and $\lambda_\gamma$, taking the values 0 (longitudinal polarisation) and $\pm 1$ (transverse polarisations). The helicity amplitudes, 
$T_{\lambda_\rho \lambda_\gamma}$, extracted at HERA~\cite{Aaron:2009xp,Chekanov:2007zr} respect a strict ordering, which reflects the strong influence of small-size dipole scatterings, as argued in Ref.~\cite{Ivanov:1998gk}:
\begin{equation}
T_{-11} \ll T_{01} \ll T_{10} \ll T_{11} \ll T_{00} \; .
\end{equation}
Experimental data have been selected in distinct ranges of the photon virtuality, $Q^2$, and of the photon-proton centre-of-mass energy, $W$. Following the cuts used by H1, one has:
\begin{equation}
2.5 \; {\rm GeV}^2 < Q^2 <60 \; {\rm GeV}^2
\end{equation}
and
\begin{equation}
35 \; {\rm GeV} < W < 180 \; {\rm GeV} \; . 
\end{equation}

\subsection{Helicity amplitudes in high-energy factorisation}
\label{leptoproduction}

In the high-energy region, $s\equiv W^2\gg Q^2\gg\Lambda_{\rm QCD}^2$, which
leads to small $x=Q^2/W^2$, the forward helicity amplitude for the $\rho$-meson leptoproduction can be presented, in high-energy factorisation, as
the convolution between the impact factor, $\Phi^{\gamma^*(\lambda_\gamma) \to \rho(\lambda_\rho)}(\kappa^2,Q^2)$, describing the $\gamma^* \to \rho$ transition, and the UGD, ${\cal F}(x,\kappa^2)$:
\begin{equation}
 \label{amplitude}
 T_{\lambda_\rho\lambda_\gamma}(s,Q^2) = \frac{is}{(2\pi)^2}\int \frac{\drv^2\kappa}
 {(\kappa^2)^2}\Phi^{\gamma^*(\lambda_\gamma)\rightarrow\rho(\lambda_\rho)}(\kappa^2,Q^2)
 {\cal F}(x,\kappa^2),\quad x=\frac{Q^2}{s}\,.
\end{equation}

Analytic formulae for both the longitudinal and the transverse impact factor can be found in Eqs.~(33) and ~(38) of Ref.~\cite{Anikin:2009bf}.
\emph{Inter alia}, a twist-2 distribution amplitude (DA)~\cite{Ball:1998sk} enters the expression of the longitudinal impact factor, whereas genuine twist-3 or Wandzura--Wilczek (WW) DAs~\cite{Ball:1998sk, Anikin:2011sa} are used in the transverse case.

We will make extensive use of the WW approximation, relaxing it through the inclusion of the genuine terms just in the study of systematic effects. We will employ the \emph{asymptotic} expression for the twist-2 DA (for further details see Sect. 2.2 of Ref.~\cite{Bolognino:2018rhb}).

\subsection{Models for the unintegrated gluon distribution}
\label{models}

Pursuing the goal to investigate and compare different approaches, without the ambition of a comprehensive treatment, six models for the UGD have been selected. We refer to the original works for details on the derivation of each parametrisation and limit ourselves to giving here just the analytic expression of the UGD used in our numerical study.

\begin{itemize}

 \item \textbf{An $x$-independent model (ABIPSW)}

This simple, $x$-independent model~\cite{Anikin:2011sa} purely coincides with the proton impact factor:
\begin{equation}
 {\cal F}(x,\kappa^2)= \frac{A}{(2\pi)^2\,M^2}
 \left[\frac{\kappa^2}{\kappa^2+M^2}\right]\,,
\end{equation}
where $M$ is a characteristic soft scale. Since the main observable is a ratio of amplitudes, the normalisation factor $A$ is irrelevant.

 \item \textbf{Derivative of the gluon PDF momentum}

The definition given right below,
\begin{equation}
 \label{xgluon}
 {\cal F}(x, \kappa^2) = \frac{\drv x g(x, \kappa^2)}{\drv\ln \kappa^2} \;,
\end{equation}
reflects the obvious condition that, when integrated over $\kappa^2$ up to the factorisation scale squared, $\mu_F^2$, the UGD must be related to the standard gluon density, $g(x, \mu_F^2)$.

 \item \textbf{Ivanov--Nikolaev (IN)}

The IN model, proposed in Ref.~\cite{Ivanov:2000cm}, is suited to probe different regions of the transverse momentum. While, in the high-$\kappa$ range, a collinear gluon PDF is employed, a peculiar Ansatz for the description at small $\kappa^2$ values is made~\cite{Nikolaev:1994cd}, which describes the colour gauge invariance constraints on the radiation of soft gluons by colour singlet targets. The gluon density at small $\kappa^2$ is supplemented by a non-perturbative soft component, in agreement to the colour-dipole phenomenology.
The analytic expression for this UGD is:
\begin{equation}
 {\cal F}(x,\kappa^2)= {\cal F}^{(B)}_{s}(x,\kappa^2) 
 {\kappa_{s}^2 \over 
  \kappa^2 +\kappa_{s}^2} + {\cal F}_{h}(x,\kappa^2) 
 {\kappa^2 \over 
  \kappa^2 +\kappa_{h}^2}\,.
 \label{eq:4.7}
\end{equation}
 For a complete discussion on parameters and expressions of both the soft ($s$) and the hard ($h$) terms, see Ref.~\cite{Ivanov:2000cm}.

 \item \textbf{Hentschinski--Sabio Vera--Salas (HSS)}

The HSS parametrisation, formerly employed in the analysis of DIS structure functions~\cite{Hentschinski:2012kr}, encompasses the standard definition of the UGD in the BFKL framework, given as the convolution between the gluon Green's function and a leading-order proton impact factor. This model has been adopted for the investigation of the single-bottom quark production at LHC~\cite{Chachamis:2015ona}, for the photoproduction of $J/\Psi$ and $\Upsilon$ mesons~\cite{Bautista:2016xnp} and, quite recently, for the forward Drell--Yan reaction\footnote{Pioneering studies have been conducted in this direction~\cite{Motyka:2014lya,Brzeminski:2016lwh}, their focus lying on the twist decomposition of Drell--Yan structure functions in the dipole formalism with saturation corrections, or making use of a LLA BFKL-inspired model. In a more recent analysis~\cite{Celiberto:2018muu} the agreement between theoretical predictions and experimental data has been significantly improved by including NLA BFKL effects.}. The final formula for this UGD, given in Ref.~\cite{Chachamis:2015ona} (up to a $\kappa^2$ overall factor), reads:
\begin{equation}
 \label{HentsUGD}
 {\cal F}(x, \kappa^2; \mu_h) = \int_{-\infty}^{\infty}
 \frac{\drv\nu}{2\pi^2}\ {\cal C} \  \frac{\Gamma(\delta - i\nu -\frac{1}{2})}
 {\Gamma(\delta)}\ \left(\frac{1}{x}\right)^{\chi\left(\frac{1}{2}+i\nu\right)}
 \left(\frac{\kappa^2}{Q^2_0}\right)^{\frac{1}{2}+i\nu}
\end{equation}
\[
\times \left\{ 1 - \frac{\bar{\alpha}^2_s \beta_0 \chi_0\left(\frac{1}{2}
 +i\nu\right)}{8 N_c}\log\left(\frac{1}{x}\right)
\left[\psi\left(\delta-\frac{1}{2} - i\nu\right) + \log\frac{\kappa^2}{\mu_h^2}\right]\right\}\,,
\]
where $\beta_0=(11 N_c-2 N_f)/3$, $N_f$ the number of
active quarks,
$\bar{\alpha}_s = \alpha_s\left(\mu^2\right) N_c/\pi$,
with $\mu^2 = Q_0 \, \mu_h$, and $\chi_0(\frac{1}{2} + i\nu)$ is the leading-order eigenvalue of the BFKL
kernel. Here, $\mu_h$ is a process-typical hard scale, which can be identified
with the photon virtuality, $\sqrt{Q^2}$.
In Eq.~(\ref{HentsUGD}), $\chi(\gamma)$
is the next-to-leading order eigenvalue of the BFKL kernel, collinearly improved and employing the BLM scale-optimisation method (Sect.~2 of Ref.~\cite{Chachamis:2015ona}).
The proton impact factor is described in terms of
three parameters $Q_0$, $\delta$ and ${\cal C}$, fixed via an {\em improved} description of the photon kinematics (see Sect.~3.1 of Ref.~\cite{Bolognino:2018rhb} for further details).

 \item \textbf{Golec-Biernat--W{\"u}sthoff (GBW)}

This model originates from an effective dipole cross section $\sigma(x,r)$ for the scattering of a $q\bar{q}$ pair off a nucleon~\cite{GolecBiernat:1998js}, through a Fourier transform and reads:
\begin{equation}
 {\cal F}(x,\kappa^2)= \kappa^4 \sigma_0 \frac{R^2_0(x)}{8\pi}
 e^{\frac{-k^2 R^2_0(x)}{4}}\,.
\end{equation}
For the details and discussion of the parameters of this model, see Ref.~\cite{GolecBiernat:1998js}.

\item \textbf{Watt--Martin--Ryskin (WMR)}

The UGD model introduced in Ref.~\cite{Watt:2003mx} reads:
\begin{equation}
{\cal F}(x, \kappa^2; \mu^2) = T_g(\kappa^2,\mu^2)\,\frac{\alpha_s(\kappa^2)}
{2\pi}\,\int_x^1\! \drv z\;\left[\sum_q P_{gq}(z)\,\frac{x}{z}q\left(\frac{x}{z},
\kappa^2\right) \right.
\label{WMR_UGD}
\end{equation}
\[
 \left. + \, P_{gg}(z)\,\frac{x}{z}g\left(\frac{x}{z},\kappa^2\right)\,\Theta\left(\frac{\mu}{\mu+\kappa}-z\right)\,\right]\,,
\]
where the Sudakov-like factor $T_g(\kappa^2,\mu^2)$, whose expression is given in Ref.~\cite{Watt:2003mx}, is directly connected to the probability of evolving from the scale $\kappa$ to the scale $\mu$ without parton emission. This UGD model depends on an extra-scale $\mu$, fixed at $Q$ in this work.

\end{itemize}

\section{Results and discussion}
\label{results}

\begin{figure}[tb]
\centering

\includegraphics[scale=0.27,clip]{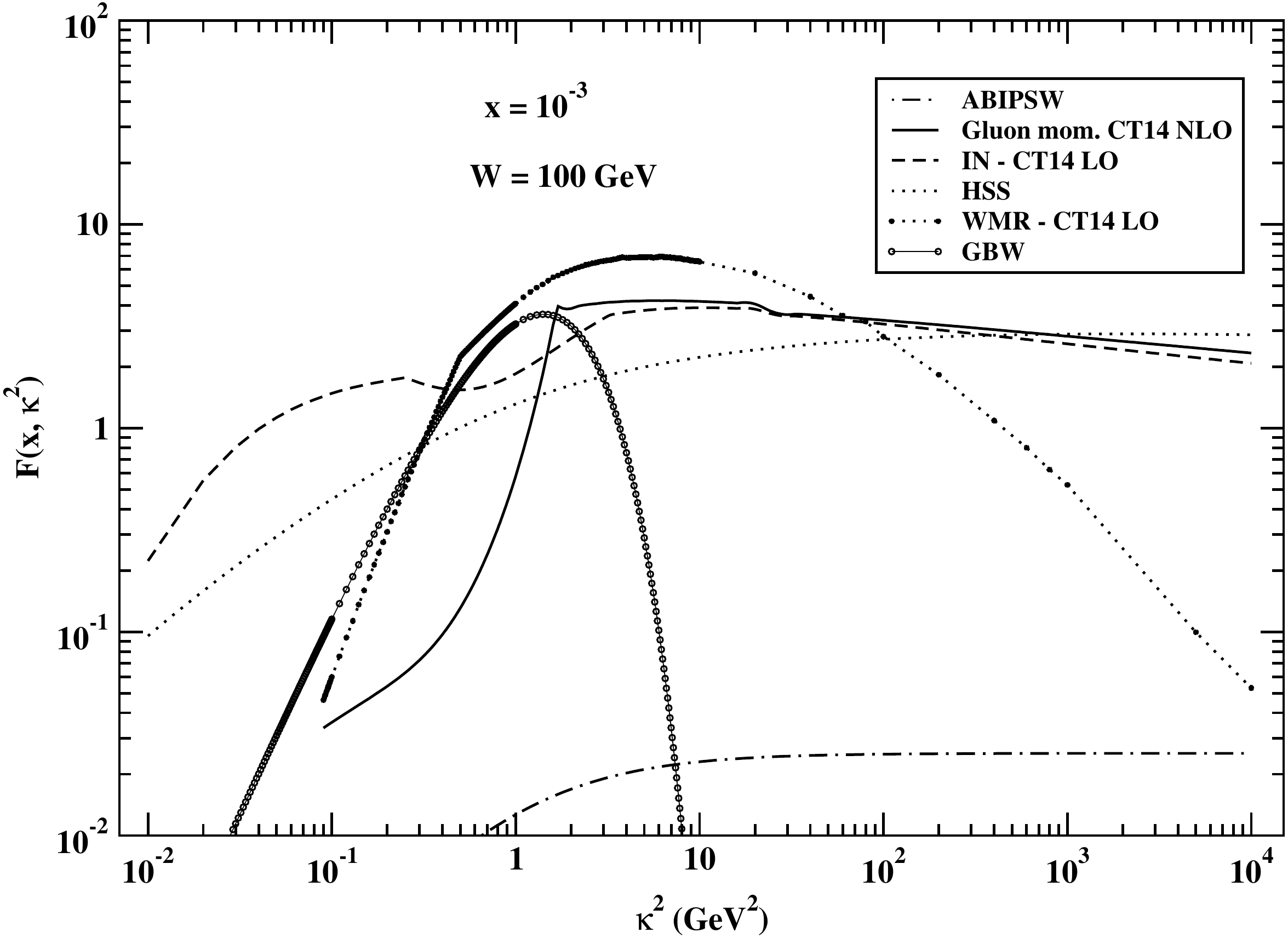}
\hspace{0.25cm}
\includegraphics[scale=0.27,clip]{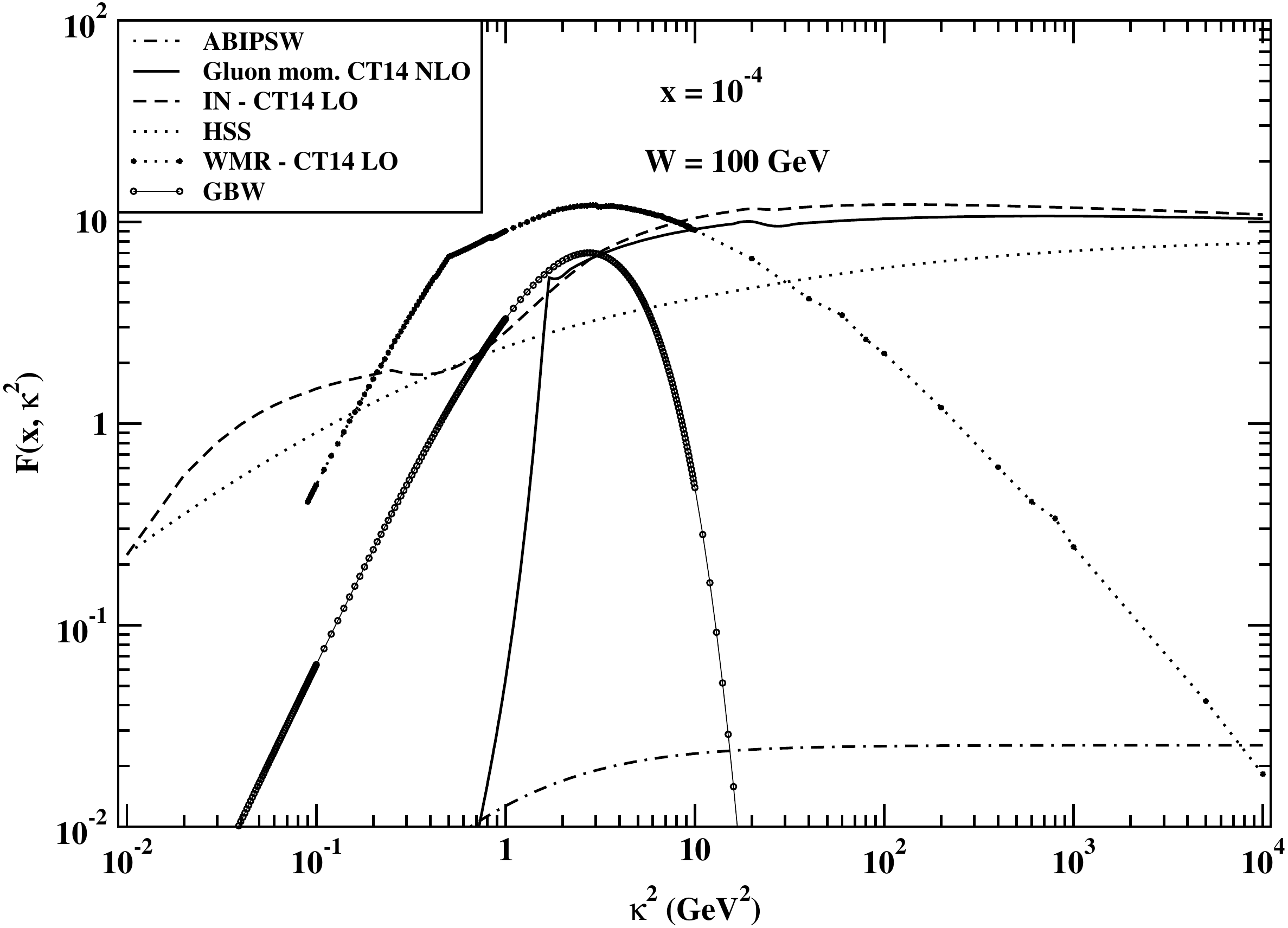}

\caption{$\kappa^2$-behaviour of all the considered UGD models for $x = 10^{-3}, 10^{-4}$.}
\label{fig:UGDs_vs_k2}
\end{figure}

\begin{figure}[tb]
\centering

\includegraphics[scale=0.27,clip]{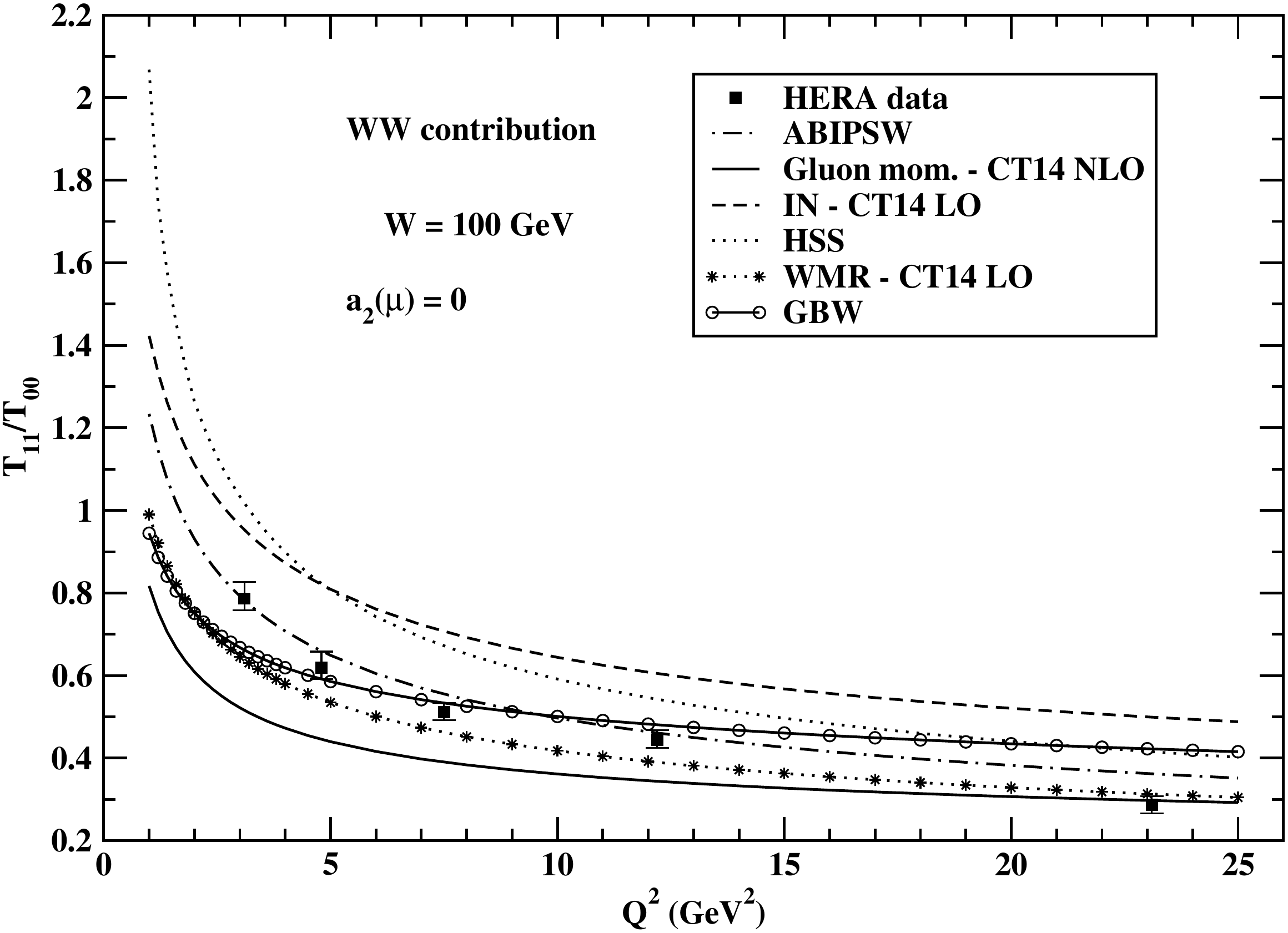}

\caption{Helicity-amplitude ratio, $T_{11}/T_{00}$, as funcion of $Q^2$ for all the considered UGD models for $W = 100$ GeV.}
\label{fig:ratio_all}
\end{figure}

\begin{figure}[tb]
\centering

\includegraphics[scale=0.27,clip]{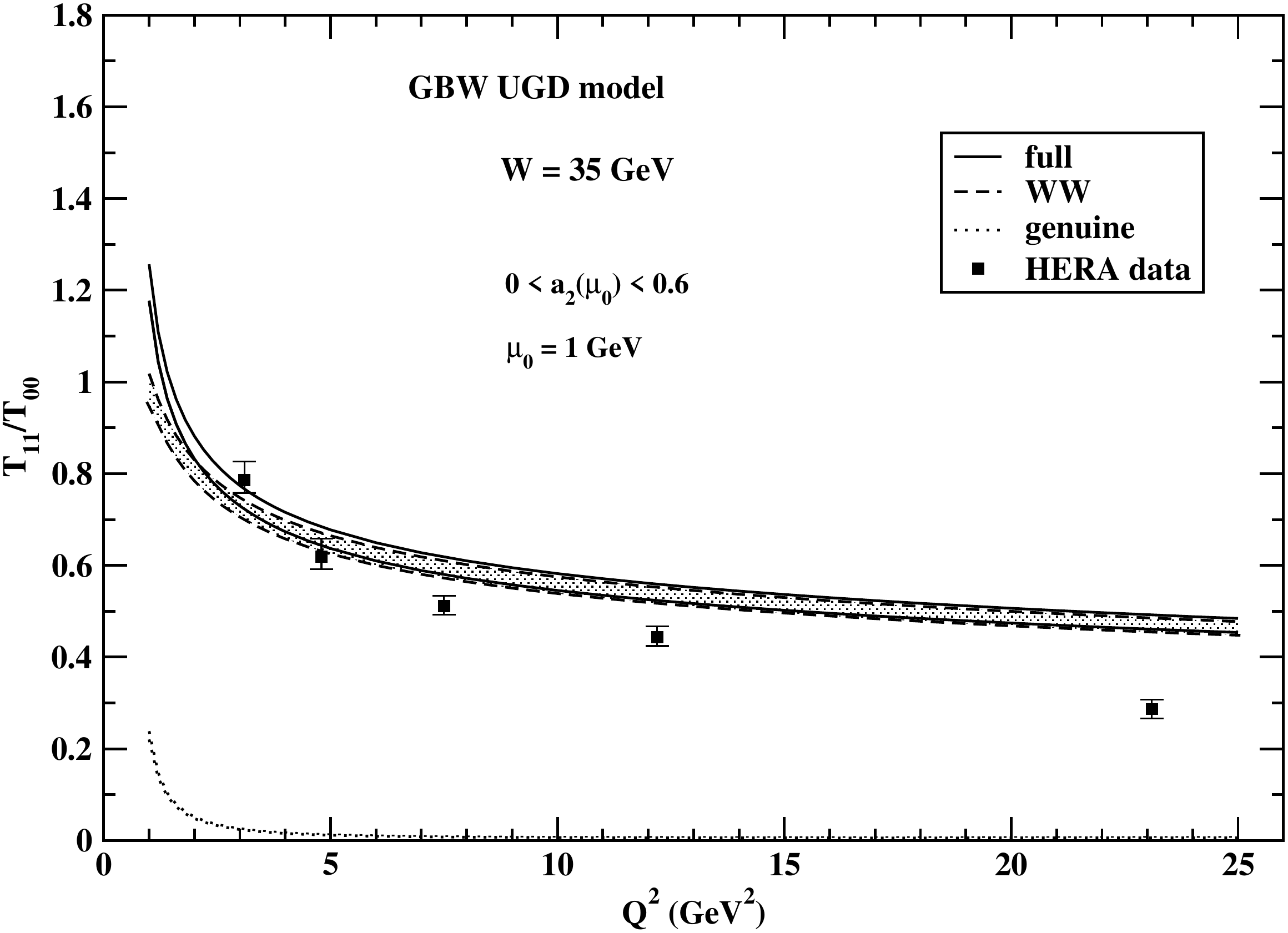}
\hspace{0.25cm}
\includegraphics[scale=0.27,clip]{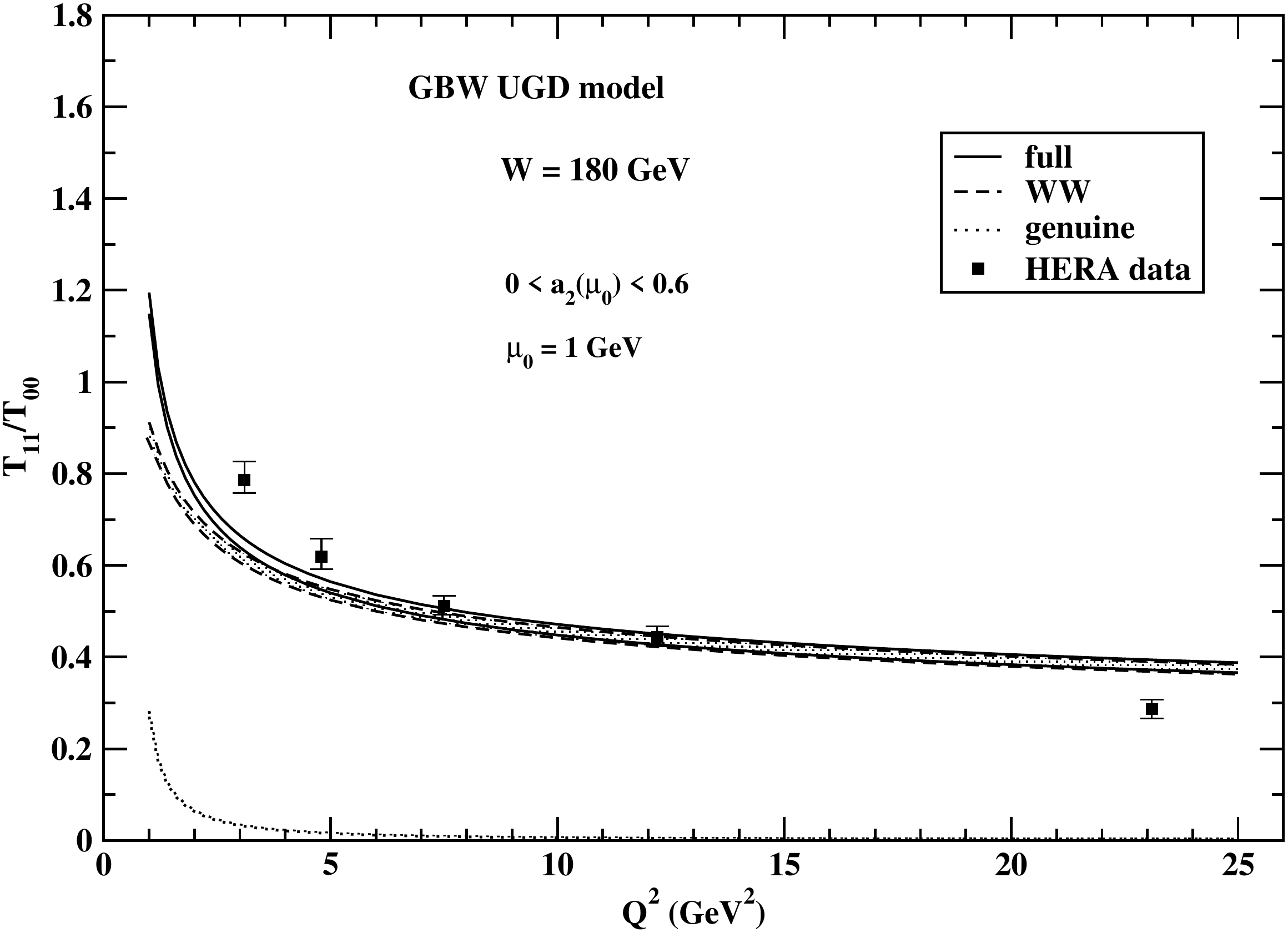}

\caption{Helicity-amplitude ratio, $T_{11}/T_{00}$, for
the GBW UGD model at $W = 35$ (left) and 180~GeV (right). Full, WW and genuine contributions are shown. Uncertainty bands are obtained by letting $a_2(\mu_0 = 1 \mbox{ GeV})$ be between 0 and 0.6.}
\label{fig:ratio_GBW_evolved}
\end{figure}

\begin{figure}[tb]
\centering

\includegraphics[scale=0.27,clip]{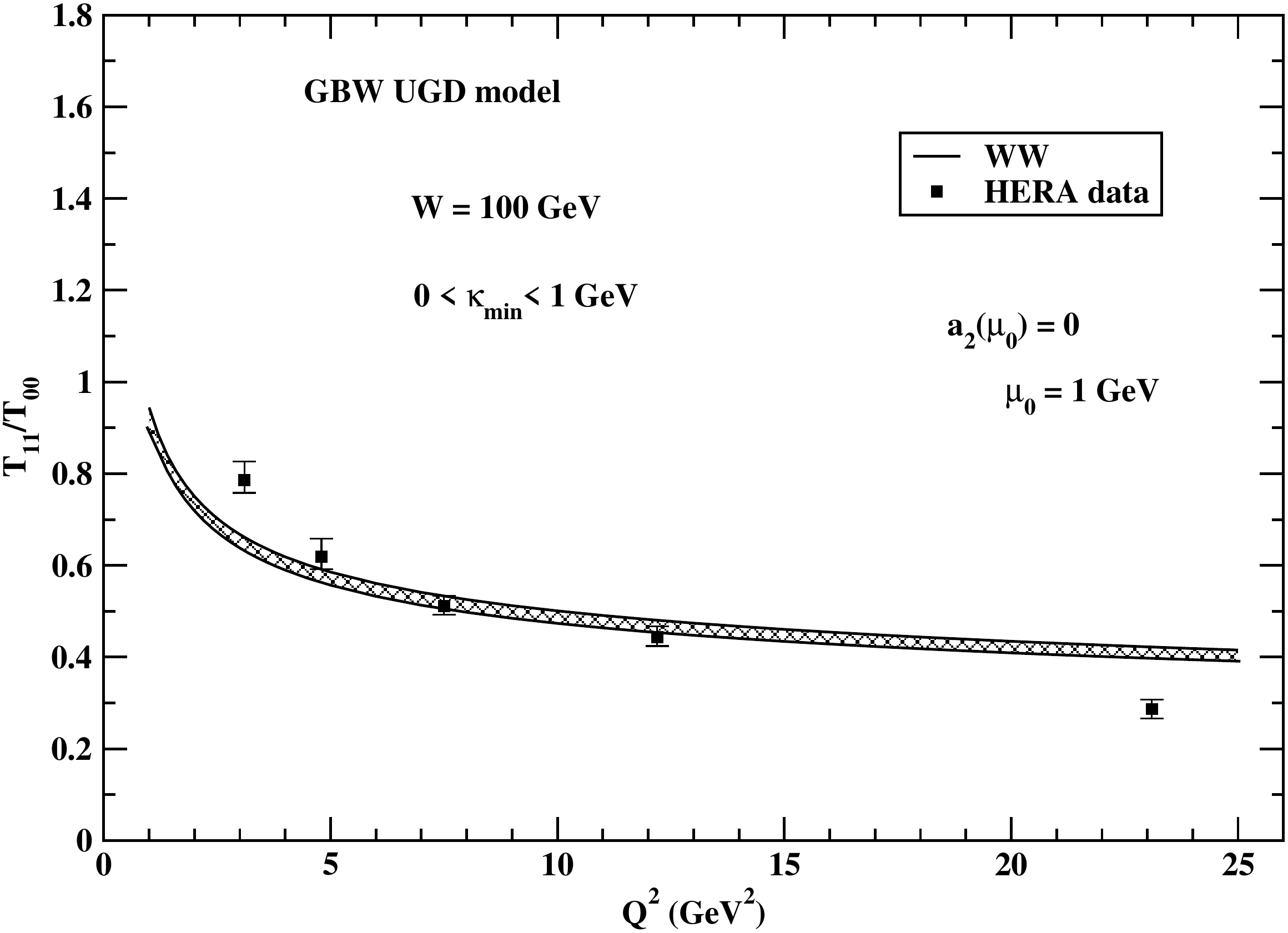}

\caption{Helicity-amplitude ratio, $T_{11}/T_{00}$, for
  the GBW UGD model at $W = 100$~GeV. The shaded band gives the effect of a lower cutoff in the $\kappa$-integration, taken in range between 0 and 1~GeV.}
\label{fig:ratio_GBW_kmin}
\end{figure}

We show the behaviour of our predictions for the helicity-amplitude ratio, $T_{11}/T_{00}$, as function of the photon virtuality, $Q^2$, and for the six different UGD models introduced in Sect.~\ref{models}, comparing them with HERA data.

First, we present and compare the $\kappa^2$-dependence of our UGD
models, for two different values of  the longitudinal momentum fraction, $x = 10^{-3}, 10^{-4}$. The different patterns in the $\kappa^2$-shape (see Fig.~\ref{fig:UGDs_vs_k2}) fairly reflect the distinct approaches whence each UGD descends.

The $Q^2$-dependence of $T_{11}/T_{00}$
for all six models at $W = 100$~GeV is then shown, together with experimental data, in Fig.~\ref{fig:ratio_all}.
Theoretical predictions are spread over a wide range, thus supporting our fundamental assertion that the $T_{11}/T_{00}$ ratio can definitely be used to constrain the $\kappa$-dependence of the UGD. 
None of the models is in agreement with data over the whole range of $Q^2$ range, whereas the $x$-independent ABIPSW parametrisation and the GBW one seem to better match the central region of $Q^2$.

In order to calibrate the effect of the approximations made in the DAs, we present results (Fig.~\ref{fig:ratio_GBW_evolved}) for $T_{11}/T_{00}$ with just one model, namely the GBW one, with the centre-of-mass energy taken at the boundaries of its interval, $W = 35, 180$ GeV, and vary the $a_2(\mu_0=1 \ {\rm GeV})$ DA parameter in a range between 0 and 0.6, correctly implementing its evolution. Furthermore, we relax the WW approximation in $T_{11}$ by accounting  also for the genuine twist-3 terms. 

Finally, we check the stability of $T_{11}/T_{00}$ under the $\kappa$ lower cut-off, keeping it in the range: 0 GeV $< \kappa_{\rm min} < 1$~GeV. In Fig.~\ref{fig:ratio_GBW_kmin} we show the result of this test for the GBW model at $W = 100$~GeV, coming out with a clear indication on the fact that the region of small values of $\kappa$ gives only a minor or negligible contribution.

\section{Conclusions}
\label{conclusions}

We have proposed the helicity amplitudes for the exclusive $\rho$-meson leptoproduction at HERA, and in possible future lepton-proton colliding-beam machines,
as an interesting and suitable probe of models of the UGD in the proton.

Theoretical reasons, backed up by accurate numerical
analyses, have been given to support our claim that both the transverse case and the longitudinal one are dominated by the kinematic region where small-size colour dipoles scatter off the proton.

In addition, we have proved that the use of distinct parametrisations for the UGD gives rise to very sparse predictions for the ratio of the transverse to longitudinal forward-scattering amplitude ratio, $T_{11}/T_{00}$.

Further tests of models for the UGD as well as the simultaneous extraction of new ones from different production channels are strongly recommendend and encouraged in the next future.  

\section*{Acknowledgments}

The author acknowledges partial support from the Italian Foundation ``Angelo Della Riccia'' and from the Italian Ministry of Education, Universities and Research under the FARE grant ``3DGLUE'' (n. R16XKPHL3N). 
The author thanks A.D. Bolognino, D.Yu. Ivanov and A. Papa for collaboration and stimulating discussions on the subject of this work.


\begin{thebibliography}{0}

\bibitem{Gribov:1984tu}
  L.V.~Gribov, E.M.~Levin, M.G.~Ryskin,
  Phys.\ Rept.\  {\bf 100} (1983) 1.

\bibitem{Celiberto:2017ius}
  F.G.~Celiberto, PhD thesis,
  arXiv:1707.04315 [hep-ph].

\bibitem{BFKL}
V.S.~Fadin, E.A.~Kuraev, L.N.~Lipatov, Phys. Lett. {\bf B60} (1975) 50;
E.A.~Kuraev, L.N.~Lipatov, V.S.~Fadin, Zh. Eksp. Teor. Fiz. {\bf 71} (1976)
840 [Sov. Phys. JETP {\bf 44} (1976) 443]; {\bf 72} (1977) 377 [{\bf 45} (1977)
199];
Ya.Ya.~Balitskii, L.N.~Lipatov, Sov. J. Nucl. Phys. {\bf 28} (1978) 822.

\bibitem{Ivanov:2004pp}
  D.Yu.~Ivanov, M.I.~Kotsky, A.~Papa,
  Eur.\ Phys.\ J.\ C {\bf 38} (2004) 195
  [hep-ph/0405297].

\bibitem{Ivanov:2005gn}
  D.Yu.~Ivanov, A.~Papa,
  Nucl.\ Phys.\ B {\bf 732} (2006) 183
  [hep-ph/0508162].

\bibitem{Ivanov:2006gt}
  D.Yu.~Ivanov, A.~Papa,
  Eur.\ Phys.\ J.\ C {\bf 49} (2007) 947
  [hep-ph/0610042].

\bibitem{Enberg:2005eq}
  R.~Enberg, B.~Pire, L.~Szymanowski, S.~Wallon,
  Eur.\ Phys.\ J.\ C {\bf 45} (2006) 759
   Erratum: [Eur.\ Phys.\ J.\ C {\bf 51} (2007) 1015]
  [hep-ph/0508134].

\bibitem{Mueller:1986ey}
A.H.~Mueller, H.~Navelet, 
Nucl. Phys. B \textbf{282}, 727 (1987).

\bibitem{Colferai:2010wu}
  D.~Colferai, F.~Schwennsen, L.~Szymanowski, S.~Wallon,
  JHEP {\bf 1012} (2010) 026
  [arXiv:1002.1365 [hep-ph]].

\bibitem{Caporale:2012ih}
  F.~Caporale, D.Yu.~Ivanov, B.~Murdaca, A.~Papa,
  Nucl.\ Phys.\ B {\bf 877} (2013) 73
  [arXiv:1211.7225 [hep-ph]].

\bibitem{Ducloue:2013wmi}
  B.~Duclou\'e, L.~Szymanowski, S.~Wallon,
  JHEP {\bf 1305} (2013) 096
  [arXiv:1302.7012 [hep-ph]].

\bibitem{Ducloue:2013bva}
  B.~Duclou\'e, L.~Szymanowski, S.~Wallon,
  Phys.\ Rev.\ Lett.\  {\bf 112} (2014) 082003
  [arXiv:1309.3229 [hep-ph]].
  
\bibitem{Caporale:2013uva}
  F.~Caporale, B.~Murdaca, A.~Sabio Vera, C.~Salas,
  Nucl.\ Phys.\ B {\bf 875} (2013) 134
  [arXiv:1305.4620 [hep-ph]].

\bibitem{Ducloue:2014koa}
  B.~Duclou\'e, L.~Szymanowski, S.~Wallon,
  Phys.\ Lett.\ B {\bf 738} (2014) 311
  [arXiv:1407.6593 [hep-ph]].

\bibitem{Caporale:2014gpa}
  F.~Caporale, D.Yu.~Ivanov, B.~Murdaca, A.~Papa,
  Eur.\ Phys.\ J.\ C {\bf 74}, no. 10, 3084 (2014)
  [Eur.\ Phys.\ J.\ C {\bf 75}, no. 11, 535 (2015)]
  [arXiv:1407.8431 [hep-ph]].

\bibitem{Ducloue:2015jba}
  B.~Duclou\'e, L.~Szymanowski, S.~Wallon,
  Phys.\ Rev.\ D {\bf 92} (2015) no.7,  076002
  [arXiv:1507.04735 [hep-ph]].

\bibitem{Caporale:2015uva}
  F.~Caporale, D.Yu.~Ivanov, B.~Murdaca, A.~Papa,
  Phys.\ Rev.\ D {\bf 91} (2015) no.11,  114009
  [arXiv:1504.06471 [hep-ph]].
  
\bibitem{Celiberto:2015yba}
  F.G.~Celiberto, D.Yu.~Ivanov, B.~Murdaca, A.~Papa,
  Eur.\ Phys.\ J.\ C {\bf 75} (2015) no.6,  292
  [arXiv:1504.08233 [hep-ph]].

\bibitem{Celiberto:2015mpa}
  F.G.~Celiberto, D.Yu.~Ivanov, B.~Murdaca, A.~Papa,
  Acta Phys.\ Polon.\ Supp.\  {\bf 8} (2015) 935
  [arXiv:1510.01626 [hep-ph]].

\bibitem{Celiberto:2016ygs}
  F.G.~Celiberto, D.Yu.~Ivanov, B.~Murdaca, A.~Papa,
  Eur.\ Phys.\ J.\ C {\bf 76} (2016) no.4,  224
  [arXiv:1601.07847 [hep-ph]].

\bibitem{Caporale:2018qnm}
  F.~Caporale, F.G.~Celiberto, G.~Chachamis, D.~Gordo G{\'o}mez, A.~Sabio Vera,
  Nucl.\ Phys.\ B {\bf 935} (2018) 412
  [arXiv:1806.06309 [hep-ph]].

\bibitem{Ivanov:2012iv}
  D.Yu.~Ivanov, A.~Papa,
  JHEP {\bf 1207} (2012) 045
  [arXiv:1205.6068 [hep-ph]].

\bibitem{Celiberto:2016hae}
  F.G.~Celiberto, D.Yu.~Ivanov, B.~Murdaca, A.~Papa,
  Phys.\ Rev.\ D {\bf 94} (2016) no.3,  034013
  [arXiv:1604.08013 [hep-ph]].

\bibitem{Celiberto:2017ptm}
  F.G.~Celiberto, D.Yu.~Ivanov, B.~Murdaca, A.~Papa,
  Eur.\ Phys.\ J.\ C {\bf 77} (2017) no.6,  382
  [arXiv:1701.05077 [hep-ph]].

\bibitem{Caporale:2015vya}
  F.~Caporale, G.~Chachamis, B.~Murdaca, A.~Sabio Vera,
  Phys.\ Rev.\ Lett.\  {\bf 116} (2016) no.1,  012001
  [arXiv:1508.07711 [hep-ph]].

\bibitem{Caporale:2015int}
  F.~Caporale, F.G.~Celiberto, G.~Chachamis, A.~Sabio Vera,
  Eur.\ Phys.\ J.\ C {\bf 76} (2016) no.3,  165
  [arXiv:1512.03364 [hep-ph]].

\bibitem{Caporale:2016soq}
  F.~Caporale, F.G.~Celiberto, G.~Chachamis, D.~Gordo~G{\'o}mez, A.~Sabio Vera,
  Nucl.\ Phys.\ B {\bf 910} (2016) 374
  [arXiv:1603.07785 [hep-ph]].

\bibitem{Caporale:2016xku}
  F.~Caporale, F.G.~Celiberto, G.~Chachamis, D.~Gordo~G{\'o}mez, A.~Sabio Vera,
  Eur.\ Phys.\ J.\ C {\bf 77} (2017) no.1,  5
  arXiv:1606.00574 [hep-ph].

\bibitem{Celiberto:2016vhn}
  F.G.~Celiberto,
  Frascati Phys.\ Ser.\  {\bf 63} (2016) 43
  [arXiv:1606.07327 [hep-ph]].

\bibitem{Caporale:2016pqe}
  F.~Caporale, F.G.~Celiberto, G.~Chachamis, D.~Gordo G{\'o}mez, B.~Murdaca, A.~Sabio Vera,
  JCEGI 5 (2017) no.2, 47
  [arXiv:1610.04765 [hep-ph]].

\bibitem{Caporale:2016zkc}
  F.~Caporale, F.G.~Celiberto, G.~Chachamis, D.~Gordo~G{\'o}mez, A.~Sabio Vera,
  Phys.\ Rev.\ D {\bf 95} (2017) no.7,  074007
  [arXiv:1612.05428 [hep-ph]].

\bibitem{Celiberto:2017nyx}
  F.G.~Celiberto, D.Yu.~Ivanov, B.~Murdaca, A.~Papa,
  Phys.\ Lett.\ B {\bf 777} (2018) 141
  [arXiv:1709.10032 [hep-ph]].

\bibitem{Bolognino:2019yls}
  A.D.~Bolognino, F.G.~Celiberto, M.~Fucilla, D.Yu.~Ivanov, A.~Papa,
  Eur.\ Phys.\ J.\ C {\bf 79} (2019) no.11,  939
  [arXiv:1909.03068 [hep-ph]].

\bibitem{Boussarie:2017oae}
  R.~Boussarie, B.~Duclou\'e, L.~Szymanowski, S.~Wallon,
  Phys.\ Rev.\ D {\bf 97} (2018) no.1,  014008
  [arXiv:1709.01380 [hep-ph]].

\bibitem{Bolognino:2018oth}
  A.D.~Bolognino, F.G.~Celiberto, D.Yu.~Ivanov, M.M.A.~Mohammed, A.~Papa,
  Eur.\ Phys.\ J.\ C {\bf 78} (2018) no.9,  772
  [arXiv:1808.05483 [hep-ph]].

\bibitem{Bolognino:2019yqj}
  A.D.~Bolognino, F.G.~Celiberto, D.Yu.~Ivanov, M.~M.~A.~Mohammed, A.~Papa,
  Acta Phys.\ Polon.\ Supp.\  {\bf 12} (2019) no.4,  773
  [arXiv:1902.04511 [hep-ph]].

\bibitem{Golec-Biernat:2018kem}
  K.~Golec-Biernat, L.~Motyka, T.~Stebel,
  JHEP {\bf 1812} (2018) 091
  [arXiv:1811.04361 [hep-ph]].

\bibitem{Deak:2018obv}
  M.~Deak, A.~van Hameren, H.~Jung, A.~Kusina, K.~Kutak, M.~Serino,
  arXiv:1809.03854 [hep-ph].

\bibitem{Besse:2012ia}
A.~Besse, L.~Szymanowski, S.~Wallon,
Nucl.\ Phys.\ B {\bf 867} (2013) 19
[arXiv:1204.2281 [hep-ph]].
  
\bibitem{Besse:2013muy}
A.~Besse, L.~Szymanowski, S.~Wallon,
JHEP {\bf 1311} (2013) 062
[arXiv:1302.1766 [hep-ph]].

\bibitem{small_x_WG}
J.R.~Andersen {\it et al.} [Small x Collaboration],
Eur. Phys. J. C {\bf 48} (2006) 53 [hep-ph/0604189];
Eur. Phys. J. C {\bf 35} (2004) 67 [hep-ph/0312333];
B.~Andersson {\it et al.} [Small x Collaboration],
Eur. Phys. J. C {\bf 25} (2002) 77 [hep-ph/0204115].

\bibitem{Angeles-Martinez:2015sea}
  R.~Angeles-Martinez {\it et al.},
  Acta Phys.\ Polon.\ B {\bf 46} (2015) no.12,  2501
  [arXiv:1507.05267 [hep-ph]].

\bibitem{Bolognino:2018rhb}
  A.D.~Bolognino, F.G.~Celiberto, D.Yu.~Ivanov, A.~Papa,
  Eur.\ Phys.\ J.\ C {\bf 78} (2018) no.12,  1023
  [arXiv:1808.02395 [hep-ph]].

\bibitem{Bolognino:2018mlw}
  A.D.~Bolognino, F.G.~Celiberto, D.Yu.~Ivanov, A.~Papa,
  Frascati Phys.\ Ser.\  {\bf 67} (2018) 76
  arXiv:1808.02958 [hep-ph].

\bibitem{Bolognino:2019bko}
  A.D.~Bolognino, F.G.~Celiberto, D.Yu.~Ivanov, A.~Papa,
  Acta Phys.\ Polon.\ Supp.\  {\bf 12} (2019) no.4,  891
  [arXiv:1902.04520 [hep-ph]].
  
\bibitem{Aaron:2009xp}
F.D.~Aaron {\it et al.} [H1 Collaboration],
JHEP {\bf 1005} (2010) 032
[arXiv:0910.5831 [hep-ex]].

\bibitem{Chekanov:2007zr}
S.~Chekanov {\it et al.} [ZEUS Collaboration],
PMC Phys.\ A {\bf 1} (2007) 6
[arXiv:0708.1478 [hep-ex]].

\bibitem{Ivanov:1998gk}
D.Yu.~Ivanov, R.~Kirschner,
Phys.\ Rev.\ D {\bf 58} (1998) 114026
[hep-ph/9807324].

\bibitem{Anikin:2009bf}
I.V.~Anikin, D.Yu.~Ivanov, B.~Pire, L.~Szymanowski, S.~Wallon,
Nucl.\ Phys.\ B {\bf 828} (2010) 1
[arXiv:0909.4090 [hep-ph]].

\bibitem{Ball:1998sk}
P.~Ball, V.M.~Braun, Y.~Koike, K.~Tanaka,
Nucl.\ Phys.\ B {\bf 529} (1998) 323
[hep-ph/9802299].

\bibitem{Anikin:2011sa}
I.V.~Anikin, A.~Besse, D.Yu.~Ivanov, B.~Pire, L.~Szymanowski, S.~Wallon,
Phys.\ Rev.\ D {\bf 84} (2011) 054004
[arXiv:1105.1761 [hep-ph]].

\bibitem{Ivanov:2000cm}
I.P.~Ivanov, N.N.~Nikolaev,
Phys.\ Rev.\ D {\bf 65} (2002) 054004
[hep-ph/0004206].

\bibitem{Nikolaev:1994cd}
N.N.~Nikolaev, B.G.~Zakharov,
Phys.\ Lett.\ B {\bf 332} (1994) 177
[hep-ph/9403281];
Z.\ Phys.\ C {\bf 53} (1992) 331. 

\bibitem{Hentschinski:2012kr}
M.~Hentschinski, A.~Sabio Vera, C.~Salas,
Phys.\ Rev.\ Lett.\  {\bf 110} (2013) no.4,  041601
[arXiv:1209.1353 [hep-ph]].

\bibitem{Chachamis:2015ona}
G.~Chachamis, M.~De\'{a}k, M.~Hentschinski, G.~Rodrigo, A.~Sabio Vera,
JHEP {\bf 1509} (2015) 123
[arXiv:1507.05778 [hep-ph]].

\bibitem{Bautista:2016xnp}
I.~Bautista, A.~Fernandez Tellez, M.~Hentschinski,
Phys.\ Rev.\ D {\bf 94} (2016) no.5,  054002
[arXiv:1607.05203 [hep-ph]].

\bibitem{Motyka:2014lya}
  L.~Motyka, M.~Sadzikowski, T.~Stebel,
  JHEP {\bf 1505} (2015) 087
  [arXiv:1412.4675 [hep-ph]].

\bibitem{Brzeminski:2016lwh}
  D.~Brzeminski, L.~Motyka, M.~Sadzikowski, T.~Stebel,
  JHEP {\bf 1701} (2017) 005
  [arXiv:1611.04449 [hep-ph]].

\bibitem{Celiberto:2018muu}
  F.G.~Celiberto, D.~Gordo G{\'o}mez, A.~Sabio Vera,
  Phys.\ Lett.\ B {\bf 786} (2018) 201
  [arXiv:1808.09511 [hep-ph]].

\bibitem{GolecBiernat:1998js}
K.J.~Golec-Biernat, M.~W\"usthoff,
Phys.\ Rev.\ D {\bf 59} (1998) 014017
[hep-ph/9807513].

\bibitem{Watt:2003mx}
G.~Watt, A.D.~Martin, M.G.~Ryskin,
Eur.\ Phys.\ J.\ C {\bf 31} (2003) 73
[hep-ph/0306169].
  
\end{thebibliography}
\end{document}